\documentclass[twocolumn]{aastex63}

\usepackage{amsmath,amstext}
\usepackage[T1]{fontenc}
\usepackage{apjfonts} 
\usepackage[figure,figure*]{hypcap}
\usepackage[normalem]{ulem}             

\usepackage{graphicx}

\usepackage{color}

\renewcommand{\ion}[2]{#1\,{\sc #2}}

\submitjournal{ApJ}

\shorttitle{CHIANTI v.10}
\shortauthors{Del Zanna, G. et al.}

\begin{document}

\title{CHIANTI -- an atomic database for emission lines - Paper XVI:
  Version 10, further extensions}

\correspondingauthor{G. Del Zanna}
\email{gd232@cam.ac.uk}

\author[0000-0002-4125-0204]{G. Del Zanna}
\affiliation{DAMTP, Center for Mathematical Sciences, University of Cambridge, Wilberforce Road, Cambridge, CB3 0WA, UK}

\author[0000-0003-1628-6261]{K. P. Dere }
\affiliation{College of Science, George Mason University, 4400 University Drive, Fairfax, VA 22030, USA}

\author[0000-0001-9034-2925]{P. R. Young}
\affiliation{NASA Goddard Space Flight Center, Code 671,
  Greenbelt, MD 20771, USA}
\affiliation{Northumbria University, Newcastle Upon Tyne NE1 8ST, UK}

\author[0000-0002-9325-9884]{E. Landi}
\affiliation{Department of Climate, Space Sciences and Engineering, University of Michigan, Ann Arbor, MI, 48109}




\begin{abstract}
We present version 10 of the CHIANTI package. 
In this release, we provide updated atomic models for several
helium-like ions and for all the ions 
of the beryllium, carbon and magnesium isoelectronic sequences that are abundant in astrophysical plasmas.
We include rates from large-scale atomic structure and scattering calculations that are in many cases a significant improvement over the previous version, especially for the Be-like sequence, which has useful line diagnostics to measure the electron density and temperature. 
We have also added new ions and updated several of them with new atomic rates and line identifications.
Also, we have added several improvements to the IDL software, to speed up the calculations and 
to estimate the suppression of dielectronic recombination.
\end{abstract}

\keywords{atomic processes --- atomic data --- Sun: UV radiation --- Sun: X-rays, gamma rays --- Ultraviolet: general -- line synthesis}


\section{Introduction}
\label{sec:intro}

CHIANTI\footnote{www.chiantidatabase.org} 
is an atomic database and associated codes for the analysis of optically-thin
line and continuum plasma emission. 
Version 9 \citep{dere_etal:2019} of the CHIANTI database provided significant
improvements for the X-Ray satellite lines, with new atomic data and
a new method for the calculation of the line emissivities that took
into account density-dependent effects (due to the metastability
of levels in the recombining ions). Several ions were included in the
new modeling. 

In this version we significantly extend the database.
We extend the calculations of the satellite lines to new ions, 
improve atomic rates for three isoelectronic sequences
and add to  the database  several new minor ions
(those not astrophysically abundant).
We also correct a few spectral line mis-identifications.

We use  various sources of atomic data, but for the important
collisional excitation rates (CE) by electron impact we mostly rely on 
large-scale  calculations  produced by the 
UK APAP network\footnote{apap-network.org} 
\citep[see, e.g.][]{badnell_etal:2016}.
Such calculations use the advanced $R$-matrix suite of codes
but also provide energy levels and  radiative data.
Atomic data for the Be-like, Mg-like and C-like
have recently been calculated and are included here. 
These CE rates replace for several ions approximate rates previously
calculated with the distorted-wave (DW) method.
As the focus of such calculations are the CE rates, the radiative
rates for the lowest configurations are often not as accurate as those
obtained with other codes. We have therefore included for several ions A-values for the lower configurations from mainly two sets of codes. 
The first  set is mantained  by the 
COMPAS group \cite[see, e.g.][]{jonsson_etal:2017} and consists of codes for
Multiconfiguration Dirac-Hartree-Fock  (MCDHF) calculations, providing 
very accurate  transition probabilities (typically uncertainties
are estimated to be within  a few percent).
The MCDHF calculations have been using the various improvements
\citep[e.g. GRASP2K, see ][]{grasp2k:2007,grasp2k:2013}
of Ian Grant's relativistic atomic structure program GRASP
 \citep{grant_etal:1980,dyall_etal:1989}.
The second set are the 
Multiconfiguration Hartree-Fock (MCHF) calculations,
available at NIST\footnote{https://nlte.nist.gov/MCHF/periodic.html}
\citep[see, e.g.][ for a recent description of the methods]{froese_fisher_etal:2016}.

Experimental energies are generally taken from the 
NIST\footnote{https://physics.nist.gov/PhysRefData/ASD/levels\_form.html} 
compilation, although in several cases, detailed in the comments
of each file, alternative or additional identifications have been
adopted. As the GRASP2K MCDHF calculations provide
{ very accurate  wavelengths (often close to instrument's resolution),
 in several cases they have helped to track questionable or incorrect
identifications present in NIST or in CHIANTI. 
As an example on the accuracies achievable,
the \ion{Fe}{xi} wavelengths  from $n=3$ states around 200~\AA\ 
  calculated by \cite{wang_etal:2018_fe_11} are typically within
  0.2~\AA\ the observed values (see their Table~5).
The \ion{Fe}{xii} wavelengths from $n=4$ states around 80~\AA\ 
  calculated by \cite{song_etal:2020} are typically within
  0.005--0.01~\AA\ the observed values (see their Table~3).
}

The atomic data for the Be- and Mg-like sequences were benchmarked
 by \cite{delzanna:2019_eve}
against medium-resolution quiet-Sun irradiances from 60 to 1100~\AA\
measured in 2008 by the prototype Extreme Ultraviolet Variability Experiment (EVE)
on the Solar Dynamics Observatory (SDO)  \citep{woods_etal:12}.
The comparison indicated  that, at least for medium-resolution
spectral data, the CHIANTI data are now relatively complete in the
extreme ultraviolet. 
Good overall agreement between predicted and observed irradiances was found,
at a level comparable to the accuracy of the radiometric calibration  (10-30\%).

\section{New atomic data for ions in the helium isoelectronic sequence }

The ions in the helium isoelectronic sequence produce some of the most intense lines in the X-ray spectrum.  In addition, as pointed out by \citet{gabriel_jordan_he}, selected ratios of the spectral lines provide temperature and density sensitive diagnostics that are often used in the analysis of astrophysical spectra.  The helium-like ions also produce satellite lines to the hydrogen-like emission lines {by means of dielectronic recombination}.  While these are weaker than the satellite lines produced by the lithium-like ions, they have been observed in solar flares \citep{parmar_bcs, tanaka_1982, pike_1996} and in laboratory spectra \citep{rice_2014, weller_hseq_k}.

The 1s2s $^3$S$_1$ and 1s2p $^1$P$_1$, $^3$P$_2$ and $^3$P$_1$ levels give rise to strong transitions to the ground level 1s$^2$ $^1$S$_0$ that are labeled as $z$, $w$, $x$ and $y$ in standard notation \citep{gabriel_1972}.  The temperature sensitive ratio, {\it G}, is given as $( x + y + z)/w$ and the density sensitive ratio, {\it R} is given by $z/(x + y)$.  In the case of \ion{Ne}{ix}, \citet{smith_ne_9} has shown that these ratios are influenced by radiative and dielectronic recombination.  The Version 8 release of the CHIANTI database included radiative recombination into the helium-like levels through 1s8{\it l}.  In this release, the effects of dielectronic recombination are reproduced by including the configurations above the ionization potential, 2sn{\it l} (n$=2-8$, $l=0-7$) and 2pn{\it l} (n$=2-8$, $l=1-7$). 

In this release, we provide updated atomic models for the helium-like ions of C, N, O, Ne, Mg, Al, Si, S, Ar, Ca, Fe, Ni, and Zn and new atomic models for Ti and Cr.  

\subsection{Energy levels}

The energy level files in CHIANTI contain values based on both observations and theoretical calculations.  The theoretical energy levels have either been taken from the literature or calculated with the AUTOSTRUCTURE program \citep{badnell_as}.  In the past, we have relied heavily on observed energy levels provided by \citet{NIST_ASD}.  Currently, the NIST database contains energy levels based on observed spectra for the helium-like ions of C, N, O, and Ne.  These energies appear to come from a publication by \citet{kelly} and we have used these for this release.  NIST energy levels for other members of the helium isoelectronic sequence are theoretical.  The release of version 3 of the CHIANTI database \citep{dere_v3} contained observed energy levels from version 1 of the NIST database \citep{nist_v1}.  The energies in teh current version of NIST \citep{NIST_ASD} appear to be unchanged since then.  These energies have been used to create the new CHIANTI ions  \ion{Ti}{xxi} and \ion{Cr}{xxiii}.

Because of the lack of updated energies in the NIST database, we have taken the approach of relying heavily on the calculations with AUTOSTRUCTURE.  As with our work on the lithium isoelectronic sequence \citep{dere_v9}, we have used the available energies derived from measured wavelengths to tune the AUTOSTRUCTURE calculations by means of a SHFTIC file.  {The SHFTIC file is an option that can be used when performing AUTOSTRUCTURE calculations.  AUTOSTRUCTURE does not provide energy levels of sufficient accuracy to comparison with observed spectra.  The SHFTIC file can be constructed to shift the AUTOSTRUCTURE energies to energies derived from observed spectra.} 
 For levels above the ionization potential, the same has been done but with the theoretical energies of \citet{goryaev} for the 2l2l' and 2l3l' levels and the theoretical energies of Safronova as reported in \citet{kato_safronova}.  

Recently, \citet{chantler} has suggested that there is a divergence between the theoretical energy levels and the measured energy levels for the w, x, y and z lines.  They find this discrepancy begins at around a Z of 20.  \citet{indelicato} has compared several calculations of the helium-like energies of the n=2 configuration with measurements available in the literature.  He finds a discrepancy between experiment and theory that follows a Z$^4$ behaviour and becomes apparent around Z $\sim$ 20.  However \citet{malyshev_heseq} have peformed {\it ab initio} QED calculations and have found no discrepancy with the modern measurements.

Given the lack of availability of measurement in the NIST database and the potential discrepancy between experiment and theory, we have conducted a modest search for observed wavelengths for the $w$, $x$, $y$ and $z$ lines.  Wavelengths for the $w$ line are most commonly found and the $x$ line found on occasion.

\subsubsection{Observed energy levels for S XV}

Very accurate measurements of the 1s2p $^1$P$_1$ energy are provided by \citet{kubicek} and \citet{schleinkofer_1982} provide measurements of transition energies for the 1s2p $^3$P$_1$ energy.

\subsubsection{Observed energy levels for Ar XVII}

\citet{machado_ar17} has made accurate measurements of the \ion{Ar}{xvii} $w$ line using an electron cyclotron ion source (ECRIS).  These are used to provide the observed energy of the 1s2p $^1$P$_1$ level.

\subsubsection{Observed energy levels for Ca XIX}

\citet{rice_2014} performed high resolution measurements of the \ion{Ca}{xix} $w$, $x$, $y$, and $z$ lines of plasmas created at  the ALCATOR MOD-Z tokamak and these measurements have been used in the current database.

\subsubsection{Observed energy levels for Ti XXI}

\citet{payne_ti} used an EBIT source to perform accurate measures of the wavelengths of the $w$, $x$, $y$ and $z$ lines of \ion{Ti}{xxi} and these have been inserted into the current database.

\subsubsection{Observed energy levels for Cr XXIII}

\citet{beiersdorfer_1988} have measured the wavelengths of several lines of \ion{Cr}{xxiii} with the Princeton Large Torus tokamak.  These measurements correspond to the decay of the 1s2p $^1$P$_1$, 1s4p $^1$P$_1$ and 1s5p $^1$P$_1$ levels to the 1s$^2$ $^1$S$_0$ ground level.  Their experimental precision is stated to be $\Delta \lambda$ /  $\lambda$ of 1/20000.

\subsubsection{Observed energy levels for Fe XXV}
Very accurate measurements of the 1s2p $^3$P$_1$ and the 1s2p $^1$P$_1$ levels are provided by \citet{rudolph}.  These measurements are based on synchrotron illumination of highly {ionized  Fe ions in an electron beam ion trap (EBIT),  detected with a crystal monochromator}.  In addition, \citet{briand} {have provided} measurements of the wavelength of the 1s$^2$ $^1$S$_0$ - 1s2p $^3$P$_2$ transition.  \citet{beiersdorfer_1988} have provided measurements of the same transitions as they provided for \ion{Cr}{xxiii}.

\subsection{Transition Rates}

For the various transition rates, such as A-values and autoionization rates, AUTOSTRUCTURE has been utilized.  A comparison of the autoionization rates with those of \citet{goryaev} and \citet{kato_safronova} shows good agreement for the n=2 levels, moderate agreement for the n=3 levels and decreasing levels of agreement for the n=4 and n=5 levels.  The two-photon decay rate for the 1s$^2$ $^1$S$_0$ - 1s\,2s $^1$S$_0$ transition is provided by \citet{derevianko}.

\subsection{Effective Collision Strengths}

\begin{figure}[t]
\centering
\plotone{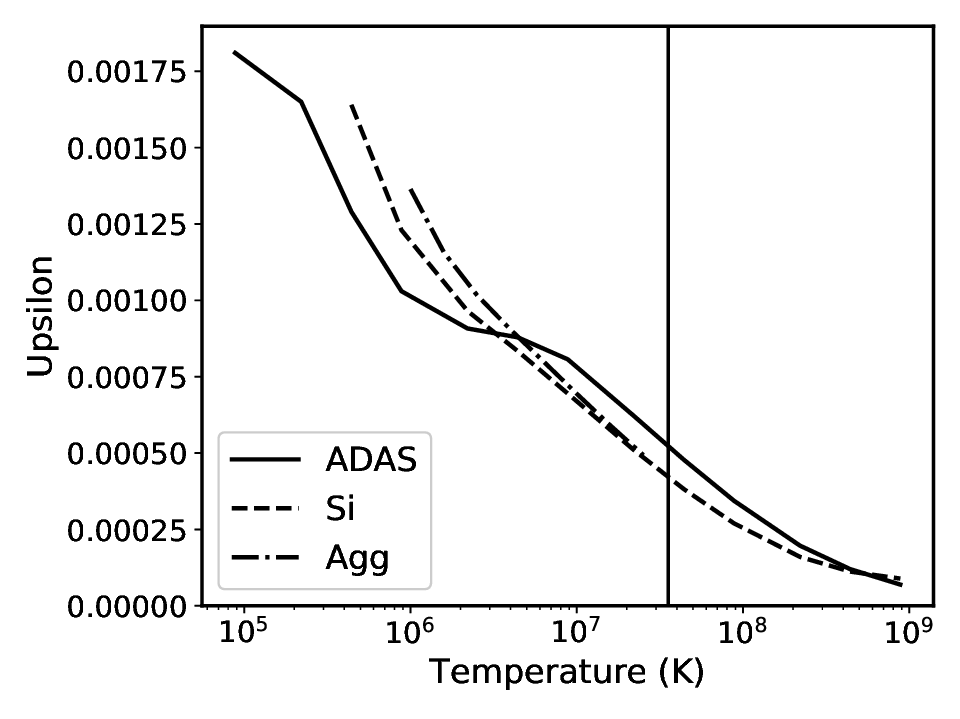}
\caption{The effective collision strengths for the 1s$^2$ $^1$S$_0$ to the 1s2s $^3$S$_1$ transition in \ion{Ti}{xxi}.  The vertical line is where the contribution function, or emissivity, for the radiative transition from the upper levels is maximum.  ADAS refers to \citet{whiteford}, Si refers to \citet{si_heseq_ups} and Agg refers to  \citet{aggarwal_ticr} 
}
\label{fig.ti21}
\end{figure}

For the pre-existing ions in the helium-like sequence, the effective collision strengths have been taken from the radiation-damped intermediate coupling frame transformation R-matrix calculations of \citet{whiteford}.   For the ions new to this release, \ion{Ti}{xxi} and \ion{Cr}{xxiii}, there are two other sets of calculations that are available, those of \citet{aggarwal_ticr} and \citet{si_heseq_ups}.  It is worthwhile to compare these effective collision strengths with those of \citet{whiteford} for the transitions from the ground level to the 1s2s and 1s2p levels for \ion{Ti}{xxi}.  The calculations of \citet{si_heseq_ups} tend to agree quite well with those of \citet{aggarwal_ticr} at about the 10\% level or better at the temperature where the calculations overlap.  However, the collisions strengths of \citet{aggarwal_ticr} do not extend to temperatures where the contribution functions for these transition maximizes ($T_{\rm max}$).  The most realistic comparison then is between the values of Si and those of Whiteford.  \citet{si_heseq_ups} employed the independent process and isolated resonances approximation using distorted waves.  The calculations were made with the Flexible Atomic Code \citep[FAC;][]{fac}. In Fig.~\ref{fig.ti21} a comparison of the three calculations for the 1s$^2$ $^1$S$_0$ - 1s2s $^3$S$_1$ collision strength is shown.  This is largely responsible for producing the $z$ line.  At $T_{\rm max}$ the Whiteford values are about 20\% higher than those of Si.  For most transitions to the 1s\,2s and 1s\,2p levels, the values of Whiteford and Si show an agreement at about the 10\% level.

For the present release, the calculations of Whiteford et al. for the ions
along the sequence have been used (and re-fitted), as made available on the UK APAP and OPEN-ADAS web sites.

\subsection{Diagnostic ratios involving the Fe XXV satellites}

\begin{figure}[t]
\centering
\plotone{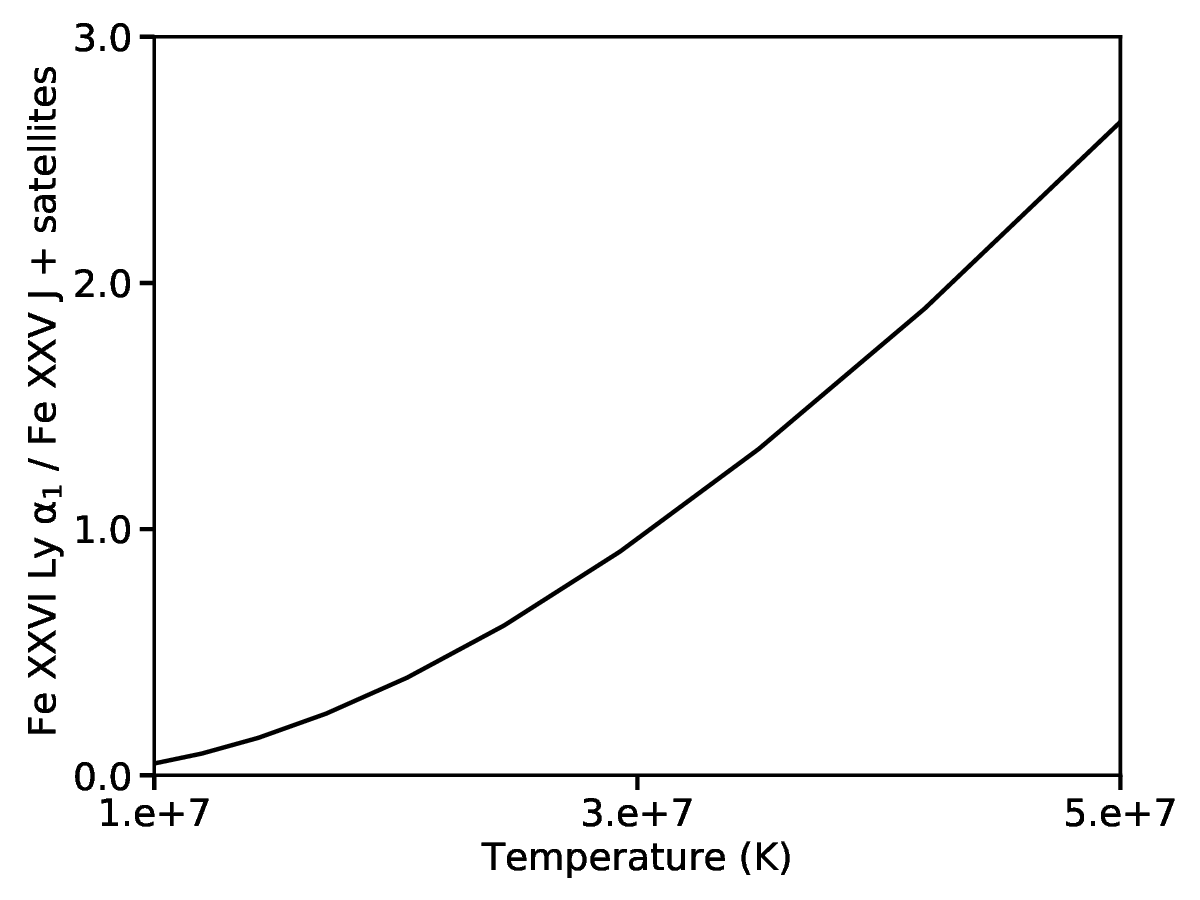}
\caption{The ratio of the Fe XXVI Ly $\alpha_1$ line to the Fe XXV J satellite and the 17 strongest nearby Fe XXV satellite lines.}
\label{fig.fe26}
\end{figure}

As noted, the helium-like J satellites to the hydrogen-like Ly$\alpha$ lines have been observed and the ratio of the two provide a temperature diagnostic for high temperature plasmas.  The J satellites are a collection of lines at somewhat longer wavelengths than the Ly$\alpha$ lines. \citet{parmar_bcs} used the calculations of \citet{dubau_fe26} for the analysis of their data.  More recently, the observations of \citet{pike_1996} were analyzed with the updated calculations of \citet{dubau_fe26}.  The new atomic data in CHIANTI allows one to calculate this ratio and the ratio is shown in Fig. \ref{fig.fe26}.  
{This consists of the ratio of the \ion{Fe}{xxvi} 1s $^2$S$_{1/2}$ - 2p $^2$P$_{3/2}$ transition at 1.7780 \AA\ to the \ion{Fe}{xxv} 1s2p $^1$P$_1$ - 2p$^2$ $^1$D$_2$ transition at 1.7920 \AA (the J satellite) and 17  other satellites at closely neighboring wavelengths. The ratio is comparable to that shown in \citet{pike_1996}, although they use the sum of the \ion{Fe}{xxv} satellite lines for values of n up to 20.  Their ratio goes to a value of about 5 at 5 $\times$ 10$^7$ K while our ratio goes to a value of about 2.7 at the same temperature.}

\section{New atomic data for the beryllium isoelectronic sequence }

It is well-known \citep[cf. the review by][]{delzanna_mason:2018}
that spectral lines from Be-like ions are prominent and provide
important density and temperature diagnostics.
Of all the isoelectronic sequences, this was the one
where the CHIANTI atomic data needed most improvements. In fact, 
for several Be-like ions, previous CHIANTI versions had either
interpolated collisional rates  \citep{keenan_etal:86}, or had sufficiently accurate data only for the $n=2$ (and sometimes $n=3$) levels. 
As previously shown in \cite{delzanna_etal:08_mg_9} for an important ion for the solar corona,
\ion{Mg}{ix}, interpolation of the rates can lead to incorrect results
such as temperatures (with differences up to 50\%). 

{
As a baseline (unless otherwise stated), we have adopted for the ions in the sequence the CE  rates
calculated by \cite{fernandez-menchero_etal:2015_be-like}
with the Breit-Pauli $R$-matrix suite of codes
and the ICFT method.}
The CI and CC expansions included atomic states up to $nl=7{\rm d}$, for  a total
of $238$ fine-structure levels. 
Good agreement with the 
previous  $R$-matrix CE rates  for \ion{Mg}{ix} 
\citep{delzanna_etal:08_mg_9} and \ion{Fe}{xxiii} \citep{chidichimo_etal:05}
were found  by \cite{fernandez-menchero_etal:2015_be-like},
but significant differences with other calculations are present.
The $n=6,7$ levels only included 
$l = s, p, d$ states, which were added to improve the  rates for the lower levels.
{The configuration expansion for the highest levels is typically unbalanced
  as energetically higher bound and continuum states are missing. Hence the
  A-values and the CE rates for some states can diverge substantially from the expected
  variation with the principal quantum number
  (see the discussion in \citealt{delzanna_etal:2020_he}). This is the case for the
  $n=6,7$ states.
  Therefore, only the 166 bound levels up to $n=5$ have been retained
  (reordering the
indexing of the original calculation),  unless otherwise
  stated. 
}

 For some ions, the  levels  above the ionization threshold  have not
 been included, as  proper treatment would require an extended two-ion model,
 including autoionisation rates.
Only transition probabilities with a branching ratio
greater than 10$^{-5}$  have been retained.

For the radiative data, we have replaced the values for the lowest
configurations with data that are generally more accurate, checking that
deviations for the strongest transitions were within 10-30\%.
{For \ion{C}{iii} and \ion{O}{v}  we adopted for the lowest 20 levels the  A-values from the
  MCHF calculations by  \cite{tachiev_cff:1999}.
  Duplicates in the MCHF data were removed, and multipolar transition probabilities 
added. 
  For the ions from Ne to Zn 
we adopted for the lowest $n=2,3$ states the  A-values from the
calculations of \cite{wang_etal:2015_be-like}, finding small
differences (10\% or so) with the APAP data.
The latter used 
a combined Configuration Interaction and Many-body Perturbation Theory Approach (MBPT).
}

Comparisons with the earlier datasets show significant changes (20--30\%)
in the line intensities for many ions,  even for the strong transitions. 
Experimental energies are taken from the NIST database, except where noted below.

\subsection{\ion{C}{iii} }

{
\ion{C}{iii} is a particularly important ion as it produces strong spectral lines. }
Only the lowest 75  bound levels up to $n=5$ were retained.
Significant differences are present 
in the density- and temperature-sensitive 
ratio of the 2s 2p $^1$P-2p$^2$ $^1$S 1247.4~\AA\ line with the lines of the
1176~\AA\ multiplet (2s 2p $^3$P-2p$^2$ $^3$P), when compared to the previous model, which included
only the $n=2,3$ levels.

\subsection{\ion{N}{iv} }

\ion{N}{iv} has been the subject of extensive studies on
the collisional rates.
\cite{aggarwal_keenan:2016} performed relativistic electron scattering
calculations for this ion with the DARC code, finding
significant discrepancies with the ICFT results of 
\cite{fernandez-menchero_etal:2015_be-like}, questioning the validity of
the ICFT method. Previous doubts from the same authors were published
\citep{aggarwal_keenan:2015}, leading to a discussion on the importance of
the target structure description by \cite{be-like_rebuttal}.

\cite{fernandez-menchero_etal:2017_n_4} performed a large-scale 
electron impact excitation calculation for \ion{N}{iv}
using the B-spline R-matrix method, and compared the results with the ICFT and
the DARC ones.  The paper highlighted the importance of the target structure description
and especially the size of the close-coupling expansion, pointing out that
collisional rates to highly excited levels can be inaccurate, because of the
limitation in the expansion. 

The three sets of atomic rates obtained with the different methods
provided an opportunity to obtain some estimates on the uncertainties, as well
as their effect on the calculated emissivities of the lines.
\cite{delzanna_etal:2019_n_4} showed that for astrophysical applications
the three sets of data are equivalent, as very small (typically 10\%)
variations in the emissivities of the stronger lines  were found.

For the present model we adopt the results of the
B-spline method, as this was the largest-scale calculation, providing much
better target energies than the other methods.
Only the lowest 136 bound levels have been retained.
Experimental energies from NIST as assessed by
\cite{fernandez-menchero_etal:2017_n_4} have been included.

\subsection{\ion{O}{v} }

{
  \ion{O}{v} is another important ion, producing strong spectral lines
  widely used in solar physics}.
It is worth noting that the new model provides electron densities
from within the lines in the 760~\AA\ multiplet that are about 60\% higher than
the previous model. These lines have been resolved with e.g. the 
Solar and Heliospheric Observatory (SoHO)
Solar Ultraviolet Measurements of Emitted Radiation (SUMER, see 
\citealt{wilhelm_etal:95}), and are now observed with the Solar Orbiter 
Spectral Imaging of the Coronal Environment (SPICE)
spectrometer \citep{anderson_etal:2020}. 

The previous CHIANTI model
was quite large, as it included a full set of configurations up to $n=5$.
However, the collisional rates for most levels were calculated
with the distorted wave method, except for the lower levels
for which the results of an unpublished $R$-matrix calculation were used.

Experimental energies are taken from NIST. Only a few differences with the
values in the previous CHIANTI version are present. However, we caution that
several NIST energy values are not relative to the ground state, so
the corresponding wavelengths might not be accurate.

\subsection{\ion{Ne}{vii} }

The previous CHIANTI model for \ion{Ne}{vii} had 46 $n=2,3$ levels,
and $R$-matrix collisional rates.
      {We note that some of the NIST experimental energies
        are not relative to the ground state so are uncertain.}

      \subsection{\ion{Na}{viii}, \ion{Al}{x}, \ion{P}{xii}, \ion{Cl}{xiv}, \ion{K}{xvi},
        \ion{Ti}{xix}, \ion{Cr}{xxi}, \ion{Mn}{xxii}, \ion{Co}{xxiv}, and \ion{Zn}{xxvii}  }

{
The previous CHIANTI models for \ion{Na}{viii}, \ion{Al}{x}, \ion{P}{xii},
\ion{Cl}{xiv}, \ion{K}{xvi}, \ion{Ti}{xix}, \ion{Cr}{xxi}, \ion{Mn}{xxii}, \ion{Co}{xxiv}, and \ion{Zn}{xxvii}
had only the 10 lowest $n=2$ levels.
Collisional rates were either interpolated along the sequence
or were calculated with the DW codes (\ion{Ti}{xix}, \ion{Cr}{xxi}, \ion{Mn}{xxii}, \ion{Co}{xxiv}, and \ion{Zn}{xxvii}).
The present models are therefore a significant improvement over the previous ones.

For \ion{Al}{x}, experimental energies for most levels have been taken from
\cite{khardi_etal:1994} as the NIST compilation is not complete.
For \ion{Ti}{xix}, \ion{Ti}{xix}, \ion{Mn}{xxii}, and \ion{Co}{xxiv}
we note that several NIST energies  seem uncertain, and some are estimated values,
i.e. not measured relative to the ground state.

}

\subsection{\ion{Mg}{ix} }

\ion{Mg}{ix} is an important ion for the solar corona, as it provides 
excellent temperature diagnostics (available with the Solar Orbiter SPICE
spectrometer), see e.g. \cite{delzanna_etal:08_mg_9} and references therein.
The previous CHIANTI model ion had 108 levels, with $R$-matrix collisional rates
for the lowest 98 up to $n=4$, 
and a few  $n=5$ levels with DW rates.
 The collision rates are
close to the previous ones, so minor differences with the previous model are present.
 Experimental energies are taken from the previous CHIANTI version and
 from NIST.

\subsection{\ion{Si}{xi} }

The previous model had $R$-matrix collisional rates calculated with
a small target, only the 10 lowest $n=2$ levels, and only at intermediate
energies. The rest of the CHIANTI model contained DW collisional rates.

Experimental energies have been re-assessed as both 
 the previous CHIANTI version and the NIST values are surprisingly not complete.
 The values recommended by the extensive
 compilation of \cite{kramida_traebert:1995} have been adopted.
 These authors also provided $LSJ$ percentage composition which was key to
 identify many of the strongly spin-orbit mixed levels, where level
 assignment becomes difficult.
 They also provided uncertainty estimates for the observed energies.
 Their values are significantly different in many cases from those in
 NIST and in the previous CHIANTI version, but are generally very close to the
 theoretical energies of the new model.
 In a few cases, tentative energies have been assigned as observed,
 while in a couple of cases the \cite{kramida_traebert:1995} values could not
be assigned to a level. 

\begin{figure}[!htbp]
\centering
\includegraphics[width=.75\hsize, angle=-90]{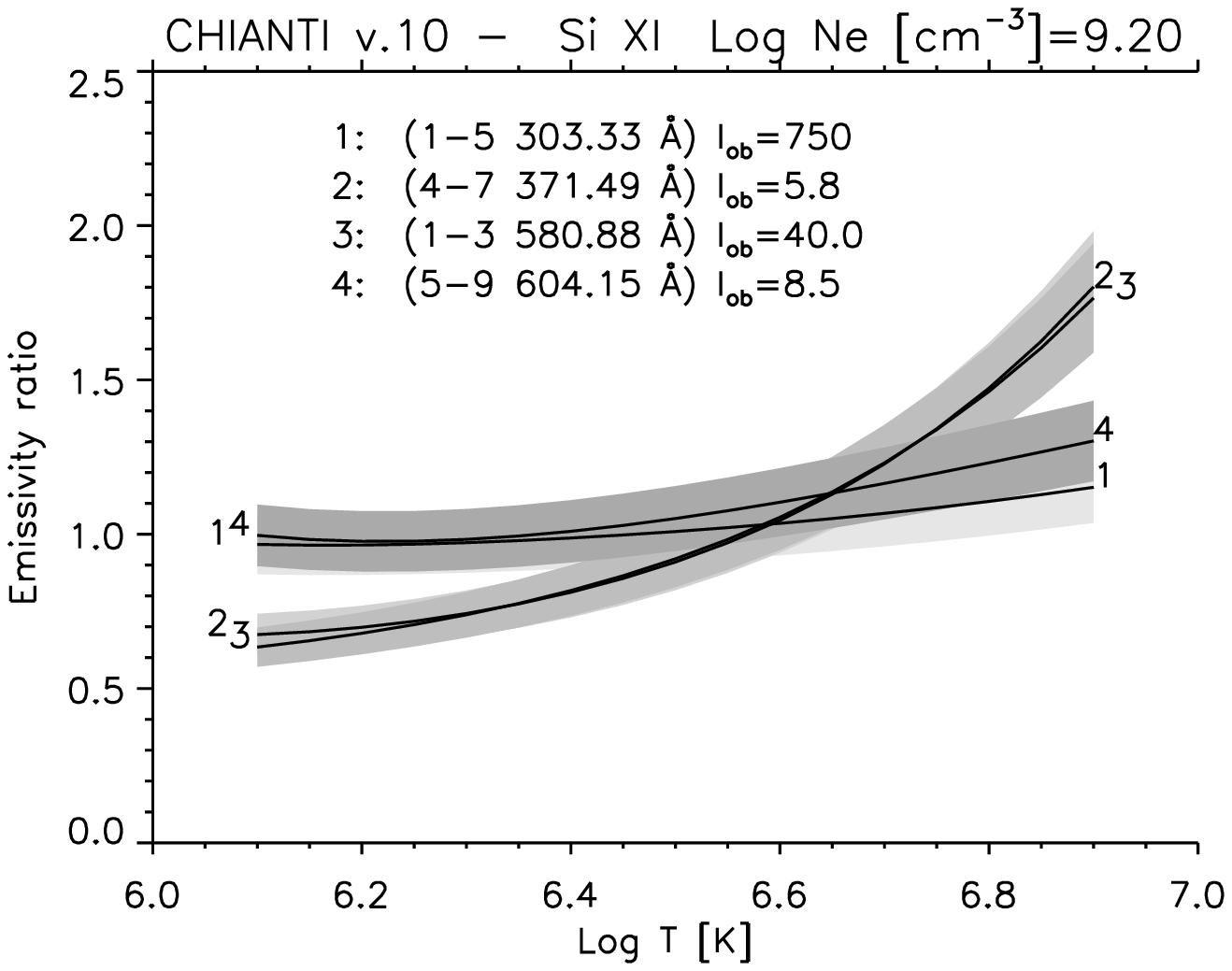}
\includegraphics[width=.75\hsize, angle=-90]{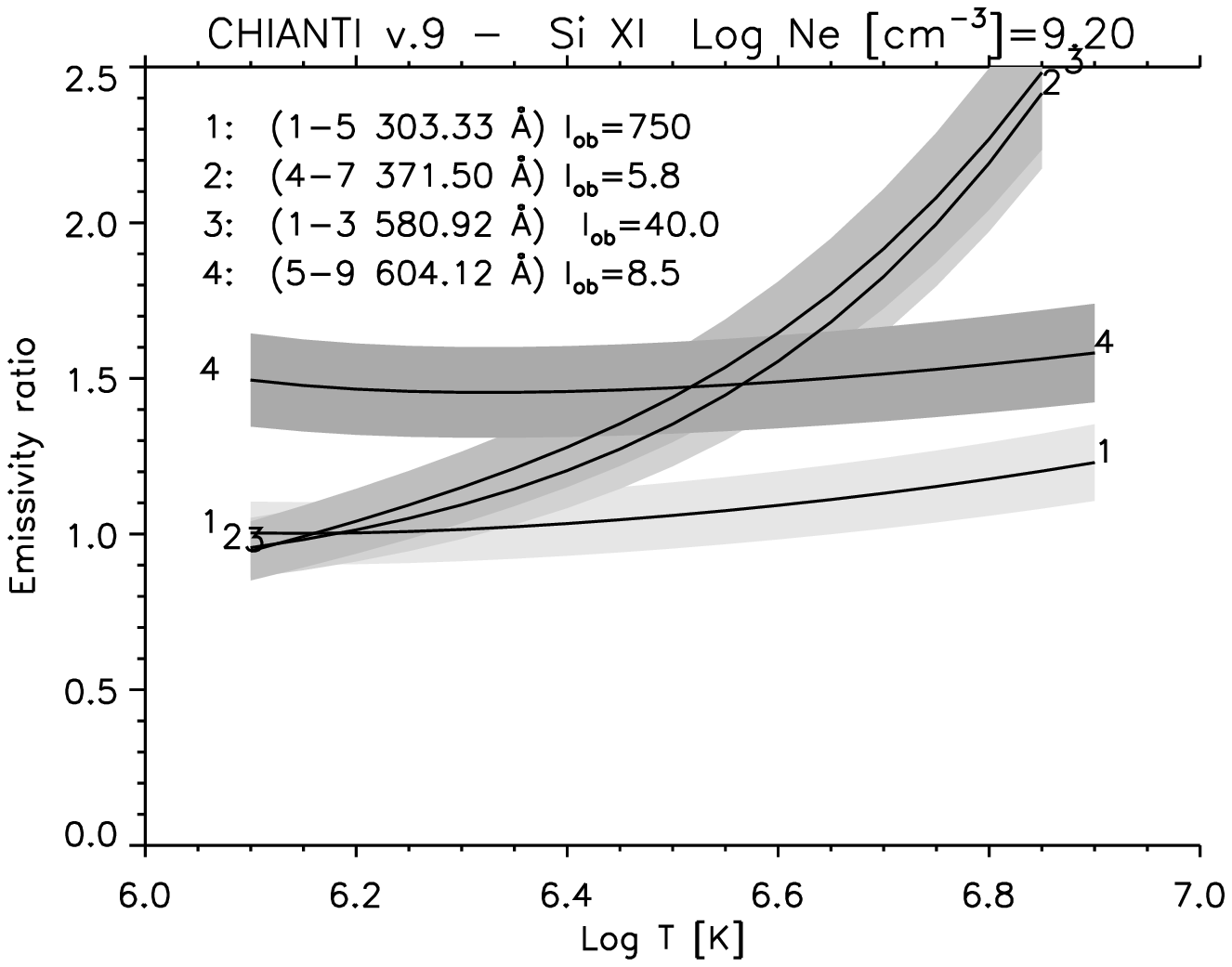}
\caption{Emissivity ratios of the main \ion{Si}{xi} lines observed by SoHO CDS in an active region,
  with the present v.10 model (above)
  and the previous one (v.9, below), as a function of the electron temperature.
  The labels indicate in parentheses the CHIANTI level indices of the lower and upper
  state, and the wavelength in \AA. 
  I$_{\rm ob}$ indicates the observed radiances in photon units.
  The shaded areas indicate the approximate uncertainty in the radiometric calibration.}
\label{fig:em1}
\end{figure}

\begin{figure}[!htbp]
\centering
\includegraphics[width=.75\hsize, angle=-90]{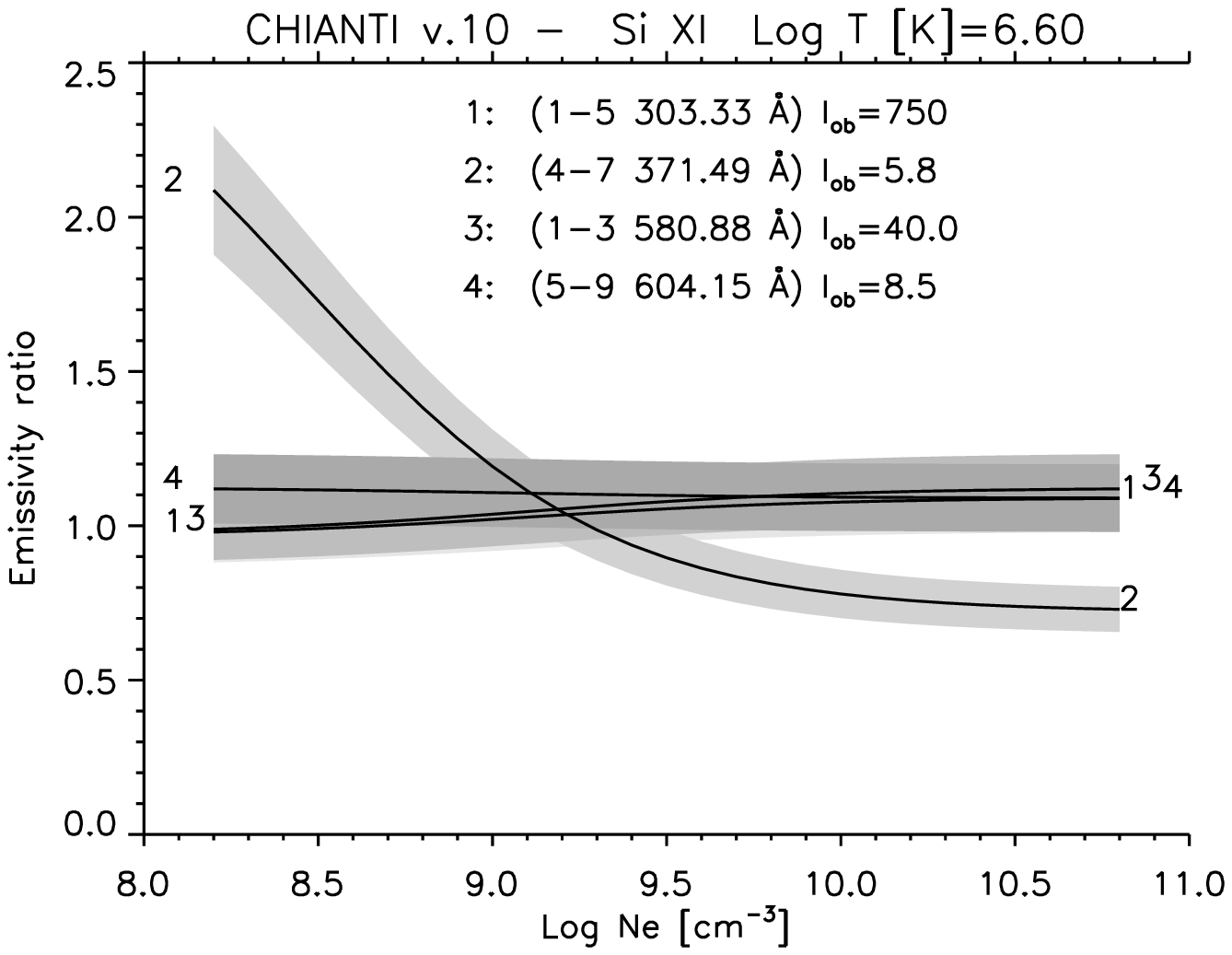}
\includegraphics[width=.75\hsize, angle=-90]{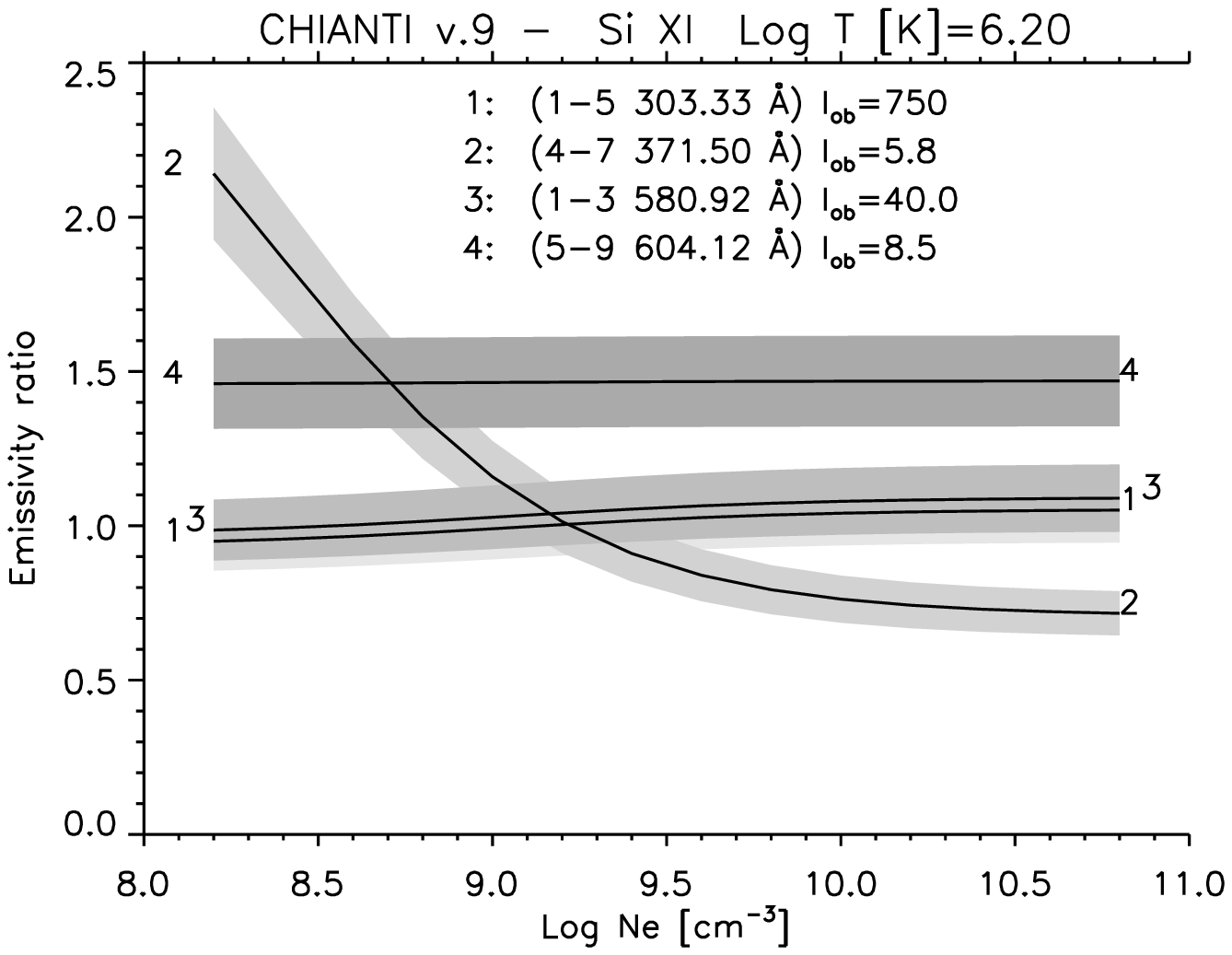}
\caption{Emissivity ratios of the same lines as in 
Fig.~\ref{fig:em1}, as a function of the electron density, with the present
model (v10, above), and the previous one (v.9, below).}
\label{fig:em2}
\end{figure}

The new model for \ion{Si}{xi} is a significant improvement
for this important ion. Long-standing problems between predicted and
observed intensities for this ion have been present for solar
active regions. The strongest lines were observed by the
SoHO CDS instrument. To illustrate the difference with the previous model,
we have considered a CDS full-spectral atlas of an active region obtained on 1997 September 25 between 18:01 and 18:56 UT 
(s9253r00). This observation was one of those considered for the  CDS in-flight calibration \citep{delzanna_thesis99,delzanna01_cdscal}.
A spatial area where the \ion{Si}{xi} lines were enhanced was selected. 
The unblended lines were selected and the comparisons between observed and 
theoretical intensities, as a function of electron density and temperature,
are presented in Figures~\ref{fig:em1},\ref{fig:em2} in the form of `emissivity ratios' 
\citep[see ][for details and examples]{delzanna_etal:2004_fe_10,delzanna_mason:2018}, which
{ are the ratios of the observed radiances $I_{\rm ob}$ of the lines
with their emissivities, as a function of the electron density $N_{\rm e}$ (or temperature $T_{\rm e})$)
for a fixed temperature (or density):
\begin{equation}
R_{ji}= { I_{\rm ob}  N_{\rm e} \lambda_{ji} \over N_j(N_{\rm e}, T_{\rm e})  \;A_{ji}} \; Const \;,
\end{equation}
\noindent
where $N_j (N_{\rm e}, T_{\rm e})$ is the
population of the upper level $j$ relative to the total
number density of the ion, $\lambda_{ji}$ is the wavelength of the transition, 
$A_{ji}$ is the spontaneous radiative transition probability,
and $Const$ is a scaling constant that is the same for all the lines within one observation.
 The value of $Const$ is chosen so that the emissivity ratios $R_{ji}$ are  near unity
where they intersect, so  the scatter of the curves around unity provides
a direct measure on the relative agreement between predicted and 
observed intensities.
}

The 371.5~\AA\ line is very weak but is strongly density-dependent.
It indicates an electron density about $\times$10$^{9.2}$ cm$^{-3}$,
with both the previous and current model. This density is typical 
of active region cores. 
The 604.1~\AA\ line is also very weak. However, with 
the present model its intensity agrees with that of the other lines.
With the previous version 9 model there is a discrepancy of 50\%.
The two main lines are the resonance line at 303.3~\AA, observed in second order,
and the intercombination line at 580.9~\AA. They are an excellent 
temperature diagnostic for active regions \citep{delzanna_mason:2018}. 
With the current model, they indicate  log $T$[K]=6.6, which is a reasonable 
value for an active region (4 MK). With the previous model, they provide a
much lower temperature:  log $T$[K]=6.2, i.e. 1.6 MK.

\subsection{\ion{S}{xiii} }

\ion{S}{xiii} is an important ion to measure the sulfur abundance in 
solar active regions.
The previous model ion had many levels (92), but CE rates
for the $n=2$ levels were interpolated along the sequence,
and the other ones were calculated with the DW approach.
Experimental energies are taken  from NIST  for the lowest
levels, but for the rest energies have been re-assessed, mostly
taking the values recommended by \cite{fawcett_hayes:1987}.

\subsection{\ion{Ar}{xv} }

The previous model ion was extended (92 levels), but CE rates
for the $n=2$ states were interpolated along the sequence,
and the other ones were calculated with the DW approach.
Experimental energies are taken  from NIST, but we note that
many of the $n=3$ values were actually estimated with quantum defects
by \cite{khardi_etal:1994}, so the
wavelengths of the decays to the lower levels are not necessarily accurate.


\subsection{\ion{Ca}{xvii} }

{The previous model ion for \ion{Ca}{xvii} was relatively accurate,
given the importance of this ion, which produces strong EUV lines}.
It had 92 levels, with 
CE rates for the $n=2$ levels calculated with the $R$-matrix
codes, and DW data for the rest of transitions.
 Experimental energies are taken from the previous CHIANTI version.

\subsection{\ion{Fe}{xxiii} }

\ion{Fe}{xxiii} is an important ion to study solar flares. 
The previous \ion{Fe}{xxiii} model had $R$-matrix CE rates for the levels up to
$n=4$, and DW rates for several $n=5$ levels.
In addition, the model had  autoionizing levels.
We note that the previous CHIANTI v.9 model had observed energies from NIST.
However, as several energies are  far from the theoretical values,
the CHIANTI v.8 experimental energies have been reinstated, with few
additional energies taken from NIST.
The data related to the autoionizing levels (producing the 
satellite lines) are the same as in version 9.

\subsection{\ion{Ni}{xxv} }

The previous model for \ion{Ni}{xxv} was relatively accurate,
as it had $R$-matrix collisional rates for the lower levels,
and DW rates for 92 levels up to $n=5$. 
Experimental energies are mostly taken  from the compilation in the
previous CHIANTI version, as it takes into account accurate measurements
not included in the NIST compilation. We note that many differences
are present between the NIST and CHIANTI energies.

\section{New atomic data for the carbon isoelectronic sequence }

The CE rates for all the ions in the C-like sequence
(from  from \ion{N}{ii} to \ion{Kr}{xxxi}), have  recently been calculated by
\cite{mao_etal:2020} 
within the UK APAP network with the Breit-Pauli $R$-matrix suite of codes
and the intermediate coupling frame transformation (ICFT) method.
They included in the configuration interaction (CI)
target and close-coupling (CC) collision expansion 
590 fine-structure levels, arising from 24 configurations
with $n\le4$ and the three configurations $2s^2 2p 5l$ with $l$=0--2.

{
The levels above the ionization threshold
have not been included for some of the ions (as described below),
as proper treatment would require
an extended two-ion model, which is left for future updates.
The A-values of the transitions among 
the ground configuration 2s$^2$ 2p$^2$ and  the first excited
2s 2p$^3$ levels have been taken from the MCHF calculations 
of \cite{tachiev_cff:2001} and the associated Tachiev (2004) 
as available online.}
Duplicates in the data were removed, and multipolar transition probabilities 
added.

Complete large-scale calculations of electron-impact excitation
with the $R$-matrix codes for C-like ions
were available in the previous CHIANTI version  only for a few ions:
Ne, S, and Fe. For most of the ions,  $R$-matrix data were only
available for the ground configuration $2s^2 2p^2$ levels, with 
DW CE rates available for the excited configurations.

Ions from the C-like sequence provide widely-used density diagnostics
\citep[see, e.g. the review by][]{delzanna_mason:2018}
thanks to the metastability of the ground configuration levels.
When benchmarking the new atomic rates, we found for many ions
only relatively minor changes (about 20\%) in the intensities of the strongest
transitions.  One exception is \ion{Ar}{xiii}, because of an
increase of about a factor of two in  the CE rates
for the forbidden transitions within the ground configuration, as shown in
\cite{mao_etal:2020}. This affects the density diagnostics.

The reason why for most other  C-like ions changes are not large
is related to the fact that the populations of the  ground configuration levels
are strongly affected not only by the CE rates within them,
but also by cascading  from higher levels. In turn,
the higher levels are mainly populated  by direct impact excitation
from the ground state via dipole-allowed transitions.
As is well-known (but also shown in \citealt{mao_etal:2020}) that
the DW CE rates are often close to the $R$-matrix ones for the strongest
allowed transitions, the result is that the populations of the
ground configuration levels with the new models are not very different,
despite some differences being present  between the new and
previous $R$-matrix CE rates  within the ground configuration.

For \ion{N}{ii} we have maintained the previous CHIANTI model as the CE rates are deemed more accurate. 

\subsection{\ion{O}{iii}}

For \ion{O}{iii} we have  retained the 177 bound levels of the 
\cite{mao_etal:2020} calculations. 
Regarding the CE rates, as pointed out by \cite{mao_etal:2020}
and \cite{morisset_etal:2020}, those obtained with 
the ICFT calculations for the 
transitions among the ground configuration states differ 
from the more accurate values such as those calculated by  \cite{2014MNRAS.441.3028S} and \cite{2017ApJ...850..147T}  
at the low temperatures typical of photoionized plasma. 
\cite{2017ApJ...850..147T} carried out a B-spline $R$-matrix calculation, but did not provide rates at sufficiently high temperatures needed for, 
e.g., solar plasma. 
\cite{2014MNRAS.441.3028S} performed a set of extended Breit-Pauli $R$-matrix calculations for transitions within the ground configuration, but only provided rates 
up to 25000 K. We have  adopted these latter rates, and used the \cite{mao_etal:2020} ones for the higher temperatures.
{In this way, we obtain accurate emissivities for the forbidden lines mostly used to study
  photoionized plasma and for the EUV lines mostly used in solar physics.}
For some transitions,
the \cite{mao_etal:2020} values at 25000 K are only 5--10\% higher than the 
\cite{2014MNRAS.441.3028S} rates, this difference being comparable with the typical uncertainty of these calculations, as discussed by \cite{2014MNRAS.441.3028S}. 
For the observed energies we used the NIST values, but
note that several are uncertain, as they were not measured
relative to the ground state.
The previous CHIANTI model had 46 levels, and a mixture of
$R$-matrix and DW data for the CE rates.

\subsection{\ion{Ne}{v} -- \ion{Ca}{xv}}

For \ion{Ne}{v}, the new model replaces one with 49 levels
and $R$-matrix CE rates.
We have  retained 304 bound levels.
In this case, nearly all the NIST experimental energies
are in good agreement with the theoretical ones.

The new model for \ion{Na}{vi} is a significant
improvement, as it replaces one with only 20 levels
and  DW CE rates (except for the ground configuration).
We have  retained 410 bound levels.

The previous model for \ion{Mg}{vii} had a mixture of scaled
$R$-matrix data for the ground configuration, and DW
rates for a total of 46 states.
We have  retained 450 bound levels.

For \ion{Al}{viii}, the present model is a significant
improvement over the previous one, which only had 20 levels 
and DW CE rates.
We have  retained 482 bound levels.
Several NIST energies were uncertain
and were not included. A few experimental energies for the
ground configuration were improved.


The previous model for \ion{Si}{ix} had 46 levels,
and a mixture of 
$R$-matrix data for the ground configuration, and DW
rates for the other states.
We have retained all 590 levels.

The new model for \ion{P}{x} is a significant improvement,
as the previous model had only 20 levels and DW CE rates.
 Several NIST energies are uncertain.

The previous model for \ion{S}{xi} was relatively extended,
with 74 levels, and a a mixture of 
$R$-matrix data for the ground configuration, and DW
rates for the other states.
For several levels we have improved the 
experimental energies over the NIST values.

The new model for \ion{Cl}{xii} is a significant improvement,
as the previous model had only 20 levels and DW CE rates.

For \ion{Ar}{xiii}, the new model produces line emissivities significantly
different than the previous one, which was limited, as it had only 15
levels and DW CE rates. A comparison with version 9
for some important lines was shown in \cite{mao_etal:2020}.
 Experimental energies are from NIST,
but we note that in many instances they are only approximate.

The new model for \ion{K}{xiv} is also a significant improvement
over the previous one, which had 20 levels and DW CE rates.

For \ion{Ca}{xv}, we have 
retained all 590 levels, and 
used the MCHF ab-initio A-values calculated by G. Tachiev (2004)
for the transitions among 
the ground configuration 2s$^2$ 2p$^2$ and  the first excited
2s 2p$^3$ levels. For these 15 states, the previous CHIANTI model had
$R$-matrix CE rates, while for the remaining 31 levels only DW data were
available. 

\subsection{\ion{Ti}{xvii} --  \ion{Zn}{xxv}}

The new models for
\ion{Ti}{xvii}, \ion{Cr}{xix}, \ion{Mn}{xx}, \ion{Co}{xxii},
\ion{Ni}{xxiii}, \ion{Zn}{xxv}
are a significant improvement
over the previous ones, which had 20 levels and DW CE rates.
The few available experimental energies are from NIST,
with the exception of \ion{Zn}{xxv}, for which we adopted the
recommended values for the lowest 20 states by
Edlen, and one level of \ion{Ni}{xxiii}.

We note that \cite{ekman_etal:2014} calculated accurate energies
and A-values with GRASP2K for the C-like ions from \ion{Ar}{xiii}
to \ion{Zn}{xxv}. If users are interested in identifying new
experimental energies, the GRASP2K values should be
considered.
We have not included the A-values from \cite{ekman_etal:2014}
as they only provided E1 transitions, and also because the
AS energies for these highly charged ions are close to
the experimental ones and the A-values are relatively accurate.

\subsection{\ion{Fe}{xxi} }

{
  \ion{Fe}{xxi} is an important ion in the sequence, given the
  large abundance of iron.  \ion{Fe}{xxi} lines are widely used in solar physics.}
The previous CHIANTI model had an extended set of CE rates calculated with the
 $R$-matrix codes by \cite{badnell_griffin:2001}. 
As shown by \cite{fernandez-menchero_etal:fe_21}, that calculation was somewhat limited
in the close-coupling expansion. 
The \cite{fernandez-menchero_etal:fe_21} calculations included the same set of configurations in the 
CI expansion as in \cite{mao_etal:2020}, but differed in some details. 
A few issues were found by \cite{mao_etal:2020} in the \cite{fernandez-menchero_etal:fe_21} data,
so we have replaced the previous model for the bound levels with the CE from Mao et al.

We have retained the experimental energies as
in version 9.
We have kept the Mao et al.\ A-values, as theoretical energies are quite close to the experimental ones.
Very little differences with the previous set of A-values is present. 
Very small differences of a few percent are found
for most of the strongest transitions, between the
previous and the new model.
The data for the autoionizing levels are the same as in the previous version.

\section{New atomic data for the Mg isoelectronic sequence }

Spectral lines from the Mg-like sequence are useful for a wide range of diagnostics
 \citep[see, e.g. the review by][]{delzanna_mason:2018}.
 Among the most important ions in the sequence for solar physics
are \ion{Fe}{xv}  and \ion{Si}{iii}. For a discussion on 
\ion{Si}{iii} see  \cite{delzanna_etal:2015_si_3}.

 For this sequence, we have adopted the collisional rates calculated by
\cite{fernandez-menchero_etal:2014_mg-like}
with the ICFT $R$-matrix  codes for all the ions from $\mathrm{Al}^{+}$ 
to $\mathrm{Zn}^{18+}$. The target includes
 a total of 283 fine-structure levels in both the 
CI target and CC collision expansions,  from the configurations  
$1\mathrm{s}^2\,2\mathrm{s^2p^6}\,3\{\mathrm{s,p,d}\}\,nl$ with
$n=4,5$, and $l=0 - 4$.
{For some ions listed below, levels above the ionization threshold
have not been included}, as proper treatment would require
an extended two-ion model, which is left for future updates.

 As in the case of the Be-like sequence, previous CHIANTI
 versions did not have atomic rates  for many minor ions, and for the less important
 ones only had a very limited set of levels and DW CE rates.
 As shown in \cite{fernandez-menchero_etal:2014_mg-like}, significant
 differences with the DW rates are present. 
 
For the radiative rates, we have mostly used the values
from the MCHF calculations by 
\cite{tachiev_cff:2002} for the lowest 26 $n=3$  levels
($3s^2,  3s 3p, 3p^2, 3s 3d, 3p 3d$).
Duplicates in the data were removed, and multipolar transition probabilities 
added. Differences with the \cite{fernandez-menchero_etal:2014_mg-like}
values were typically at most 20-30\%.
{Experimental energies are taken  from NIST unless stated otherwise.}

\subsection{\ion{Al}{ii} }

The previous CHIANTI model had 20 levels and  CE rates obtained from
various sources.
We have retained the lowest 60 bound states of the Fern{\'a}ndez-Menchero et al. 
model.
Experimental energies are taken  from NIST, but we note that a few are uncertain.
The original radiative data in \cite{fernandez-menchero_etal:2014_mg-like}
have been retained for consistency, as significant differences with the 
MCHF calculations for the higher levels are present.

\subsection{\ion{Si}{iii} }

The previous CHIANTI model had 20 levels and $R$-matrix  CE rates.
We have retained the lowest 82 bound states of the Fern{\'a}ndez-Menchero et al. 
model.
 Experimental energies are taken  from \cite{delzanna_etal:2015_si_3}
 and NIST.

\subsection{\ion{P}{iv}, \ion{Cl}{vi}, \ion{K}{viii}, \ion{Mn}{xiv}, \ion{Co}{xvi}, and \ion{Zn}{xix}  }

{\ion{P}{iv}, \ion{Cl}{vi}, \ion{K}{viii}, \ion{Mn}{xiv}, \ion{Co}{xvi}, and \ion{Zn}{xix}
 are new additions to the database. 
 For \ion{P}{iv}, \ion{Cl}{vi}, \ion{K}{viii}  we have retained,
 respectively, the lowest 117, 171, 209 bound states of the Fern{\'a}ndez-Menchero et al. 
model.

}

\subsection{\ion{S}{v}, \ion{Ar}{vii}, \ion{Ti}{xi}, and \ion{Cr}{xiii}  }

{The previous CHIANTI models for
\ion{S}{v}, \ion{Ar}{vii}, \ion{Ti}{xi}, and \ion{Cr}{xiii}
  were very limited, as they  had only  16 levels.
\ion{S}{v}, \ion{Ar}{vii}, \ion{Cr}{xiii} had DW CE rates calculated in $LS$ coupling. 
\ion{Ti}{xi} had interpolated CE rates.
For \ion{S}{v}, \ion{Ar}{vii} 
we have retained the lowest 159, 196   bound states of the Fern{\'a}ndez-Menchero et al. 
model.
Significant  differences with the previous  CE rates are present, as shown
for \ion{S}{v}  by Fern{\'a}ndez-Menchero et al.
}

\subsection{\ion{Ca}{ix} }

{The previous CHIANTI model contained 
complete set of 283 levels up to $n=5$, but DW
CE rates. }
Autoionising levels were also present.
We have retained the lowest 220  bound states of the Fern{\'a}ndez-Menchero et al. 
model.
Experimental energies are taken from the previous CHIANTI version.

\subsection{\ion{Fe}{xv} }

The previous CHIANTI model was extended, with the
complete set of 283 levels up to $n=5$, and $R$-matrix CE
rates for the $n=3$ levels, and  DW
CE rates for the higher levels.
The experimental energies of the previous version 9 
are retained. 
As mentioned in \cite{fernandez-menchero_etal:2014_mg-like}, significant
 differences with the previous $R$-matrix CE
rates are present in some cases, for this important ion.
As a follow-up,  \cite{fernandez-menchero_etal:2015} discussed in more detail
some \ion{Fe}{xv} transitions, and presented a benchmark against
solar data. 

\subsection{\ion{Ni}{xvii} }

The previous CHIANTI model was relatively extended, with the
complete set of 159 levels up to $n=5$, but DW
CE rates. 
Observed energies are taken from a variety of sources, as detailed
in the database.

\section{Other updated ions}

\subsection{\ion{Fe}{xii}}

Experimental energies of a few $n=3$ levels in  \ion{Fe}{xii} were changed in version 9 following the work of \cite{wang_etal:2018_fe_12}, which confirmed most
of the identifications by \cite{delzanna_mason:05_fe_12}. 
However, by mistake the indexing of two levels (37,38) was interchanged, so the energies for these two levels were incorrect. This has been fixed. 
The energies of a few $4d$ levels have now 
been changed, following \cite{song_etal:2020}. These authors have carried out  MCDHF calculations and compared predicted and observed wavelengths for the $n=4$ levels. 
Excellent agreement with the wavelengths of the new identifications proposed by \cite{delzanna:12_sxr1} was found. Generally, excellent agreement was also found with the wavelengths and identifications by  \cite{fawcett_etal:72}, with the exception of several $4d$ levels. After checking the original plates used by 
\cite{fawcett_etal:72}, and estimating line intensities based on the \cite{delzanna_etal:12_fe_12} calculations, \cite{song_etal:2020} recommended a few new values, which have been adopted here. 

\subsection{\ion{Fe}{xi}}

\cite{wang_etal:2018_fe_11} carried out extensive 
MCDHF calculations for the $n=3$ levels in \ion{Fe}{xi}, and compared energies with literature values. 
For the $3s^2 3p^3 3d$ levels, producing the strongest lines, they found excellent agreement (within a few hundreds of Kaysers) with all the new identifications proposed by \cite{delzanna:10_fe_11} and included in previous CHIANTI versions. 
For the $3s 3p^3 3d$ levels, the four previous tentative identifications proposed by \cite{delzanna:10_fe_11} have been revised following \cite{wang_etal:2018_fe_11}.

\subsection{\ion{Fe}{x}}

\cite{wang_etal:2020} carried out extensive 
MCDHF calculations for \ion{Fe}{x}, and assessed the 
wavelengths and identifications, as present in NIST,
the previous CHIANTI versions, and literature values.
Good general agreement with the values proposed by \cite{delzanna_etal:2004_fe_10} and
\cite{delzanna_etal:12_fe_10}, included in previous
CHIANTI versions was found, with notable discrepancies in a few NIST energies. 
On the basis of the 
\cite{wang_etal:2020} assessment, we have introduced 8 
revised experimental energies, three from 
\cite{jupen_etal:93}, one from \cite{delzanna_etal:12_fe_10} and four from NIST.
 
The energy separation of the $^4$D$_{7/2}$ and $^4$D$_{5/2}$ has been modified by using the recent measurement by 
\cite{landi_etal:2020ApJ...902...21L}, decreasing it from the previous 5.5~cm$^{-1}$ value (see the discussion in \citealt{delzanna_etal:2004_fe_10})
to 2.29~cm$^{-1}$. This change has negligible effects on the wavelengths of the  lines emitted by these levels around 257.26~\AA, but it has a critical importance for the measurement of magnetic field strength using the magnetically induced transition from the $^4$D$_{7/2}$ level to the ground (Landi et al. 2020b, in press).

\subsection{\ion{Ni}{xii}}

Regarding \ion{Ni}{xii}, we base our new model on the 
Breit-Pauli $R$-matrix calculations carried out with  the ICFT method
by \cite{delzanna_badnell:2016_ni_12}. 
The CI expansion included the complete set of atomic states up to $n=4$, but
for the CC expansions in the scattering calculation only the lowest 312 $LS$ terms were retained. This is a significant improvement over the previous calculations
(available in the previous CHIANTI version)
which included only the lowest 14 terms (up to 3d), producing 31 fine-structure levels. 
We provide only the lowest 
 589 fine-structure levels, as the higher ones are not reliable. 
 A key feature of the \cite{delzanna_badnell:2016_ni_12} calculations
 is the inclusion of semi-empirical 
 corrections in the structure and scattering calculations. This provides a 
 significant improvement in the rates. 
 
 Very few energy levels were known for this ion, the most important ones
 only relative to the 3d $^4$D$_{7/2}$ (the same situation which occurred in
 \ion{Fe}{x}, resolved by \citealt{delzanna_etal:2004_fe_10}).
 \cite{delzanna_badnell:2016_ni_12} provided a discussion of the various 
 optional line identifications, and suggested several new ones.  
The MCDHF calculations  by \cite{wang_etal:2020} were used by the same authors
to confirm  some of those identifications and suggest new ones. We adopt these latest
values in the present model, but point out that 
further laboratory studies are needed to confirm these identifications.

\subsection{Cr\,VIII}

\ion{Cr}{viii} was added to CHIANTI in version~6 \citep{2009A&A...498..915D} using data from an unpublished calculation of E.~Landi. This has been replaced with the data from \citet{2009A&A...506.1501A} giving a model with 362 fine structure levels from 10 configurations. 

Experimental energies are only available for 21 levels and have been taken from the NIST database. 
\citet{2009A&A...506.1501A} performed a number of structure calculations, and the theoretical energies used for CHIANTI are those from the GRASP calculation with Breit and QED corrections. Radiative data are also from the GRASP calculation. This data-set was filtered such that, for each atomic level, only those decay rates that were $> 10^{-5}$ of the maximum decay rate from that level were retained.

Upsilons were calculated by \citet{2009A&A...506.1501A} using the Flexible Atomic Code and were tabulated at nine temperatures from $\log\,T=5.0$ to 6.6 at 0.2~dex intervals. \ion{Cr}{viii} has 10 metastable levels plus the ground state and only those transitions from these 110 levels  were included in the CHIANTI data file. The collision dataset was assessed following the procedure described in CHIANTI Technical Report No.~4 and 69 allowed transitions were found not to tend towards their high temperature limits, with the limits generally higher than the upsilon values sometimes by as much as a factor 100. These discrepancies may stem from the different structure models used for the radiative and collisional calculations. For these 69 transitions, the high temperature limits were removed from the fits. The final CHIANTI scaled-upsilon file reproduces the original upsilon data to better than 0.2\%.

\subsection{Fe\,VII}

\ion{Fe}{vii} has not been updated since CHIANTI 5 \citep{2006ApJS..162..261L}, and the model contained only the nine levels of the ground configuration. The new model has 189 fine-structure levels of the $3p^63d^2$, $3p^53d^3$, $3p^53d4l$ ($l=s,p,d,f$) and $3p^53d5l$ ($l=s,p$) configurations.
The model was not previously updated because inconsistencies 
between line intensities calculated with the \cite{2008A&A...481..543W} model
and solar observations (using the NIST identifications) were noted, but different explanations were given: \cite{2009ApJ...707..173Y} revised some NIST 
identifications but assumed that the inconsistencies were due to problems in the 
atomic rates. \cite{delzanna:09_fe_7} provided an alternative set of identifications where
agreement with observations was obtained, pointing out that the \cite{2008A&A...481..543W} model was relatively correct, by comparison with a large-scale benchmark calculation. Later, \citet{2014ApJ...788...24T} provided a 
new large-scale scattering calculation producing similar rates to those of 
\cite{2008A&A...481..543W}. 

The new model  follows the assessment of Young et al.~(2020, submitted),
which relies on new laboratory measurements. Some inconsistencies are still present but the new model represent a significant improvement over the previous one. The experimental energies are mostly taken from the NIST database, but some new values are taken from Young et al. and \citet{2009ApJ...707..173Y}. Theoretical energies  for 134 levels are taken from  \citet{2018MNRAS.479.1260L}, while the remaining energies are from \citet{2014ApJ...788...24T}. 
We note that the \citet{2018MNRAS.479.1260L} energies show much better agreement with the experimental energies than those of \citet{2014ApJ...788...24T}, and so they yield more accurate wavelengths for transitions involving levels without experimental energies.

Radiative decay rates for allowed transitions are from \citet{2014ApJ...788...24T}. Forbidden transition data are from \citet{2018MNRAS.479.1260L}, supplemented with values from \citet{2008A&A...481..543W}, where necessary.

Collisional excitation rates are from \citet{2014ApJ...788...24T}. The CHIANTI assessment procedure identified a number of allowed transitions for which the rates did not tend towards the high temperature limit points, usually with the latter being too high. These cases likely arise because the decay rates were derived from a more elaborate structure calculation than used for the scattering calculation. They were treated by adjusting the upsilon scaling parameter to enable a smooth transition to the high temperature limit.

\subsection{O\,II}

It was brought to our attention that four A-values within the ground configuration of the N-like \ion{O}{ii} were incorrectly assigned. We have now replaced them with the correct ones, noting that this error only affected the intensities of some of the forbidden lines within the ground configuration. We have also improved the experimental energies and associated wavelengths using literature values.

\section{New additions to the database}

\subsection{Cr\,II}

The vanadium-like ion \ion{Cr}{ii} has been added to CHIANTI. \citet{2020ApJ...888...10T} provided radiative decay rates and Maxwellian-averaged collision strengths (upsilons) for 512 fine structure levels of the $3d^5$, $3d^45s$, $3d^44p$, $3d^34s^2$ and $3d^34s4p$ configurations. Experimental energies were available for 378 levels and were taken from the NIST database. We note that the agreement between the experimental energies and the \citet{2020ApJ...888...10T} calculated energies is excellent, with a mean discrepancy of 38~cm$^{-1}$. The radiative decay rate data-set was filtered such that, for each atomic level, only those decay rates that were $> 10^{-5}$ of the maximum decay rate from that level were retained.

Based on the radiative data we determined that there are 117 metastable levels (including the ground) and so the upsilon data-set was filtered so that only those transitions for which the lower level was a metastable were included. Upsilons were tabulated for 20 temperatures between $10^3$ and $10^5$~K. The assessment method described in CHIANTI Technical Report No.~4 was applied to the data-set and no problem transitions were found. The scaled upsilon data stored in CHIANTI reproduce the original data to better than 0.5\%\ for all transitions and temperatures.

\section{Population lookup tables and other improvements}

IDL software has been written to generate lookup tables for level populations for all ions in the database. The CHIANTI software can then use these tables to obtain level populations rather than compute them from the atomic data. This has significant time savings for the large ion models in CHIANTI. For example, a synthetic spectrum that would usually take about 10~minutes to compute can be obtained in less than a minute with the lookup tables.

To create the lookup tables, the software scales the level populations by one of two formulae, depending if the level is a metastable or not:
\begin{equation}\label{eq.n}
    n_i^\prime = \left\{ \begin{array}{cl} 10^{5+\alpha_i/T}n_i & \quad{\rm [regular]} \\
    10^{20 +\alpha_i/T}{n_i \over N_{\rm e} } & \quad{\rm [metastable]}\\ \end{array} \right.
\end{equation}
where $T$ is temperature, $N_{\rm e}$ is the electron number density, $n_i$ the level population computed by the IDL routine \verb|pop_solver|, and $\alpha_i$ a scaling parameter described below. The $n_i^\prime$ are computed for a temperature range $T_0$ to $T_1$ determined from the default ionization equilibrium file distributed with CHIANTI to be those temperatures for which the ionization fraction values are $\ge 10^{-8}$ of the peak value. The density range is set by the user. The temperature and density grids are set on a logarithmic scale with intervals of 0.05 and 0.2~dex, respectively. 

The advantage of tabulating the level populations rather than, say, the emissivities or contribution functions is that the data-set is much smaller. For example, the density range $\log\,N_{\rm e}=7$ to 13 yields tables with a total size of about 0.5~Gbytes.

The temperature dependence of a level population is usually given by $\exp(-1/T)$, hence the temperature factor in Eq.~\ref{eq.n}. The parameter $\alpha$ is determined from the following expression:
\begin{equation}
  \alpha = {T_0T_1 \over T_1-T_0} ( \log\,\bar{n}(T_1) - \log\,\bar{n}(T_0))
\end{equation}
where $\bar{n}(T)$ is the mean value of the level population over the density range of the lookup table.

The scaling applied to the level populations is chosen to remove the dominant temperature and density dependent terms from the tabulated quantity. When deriving populations from the lookup tables bilinear interpolation is used, and the accuracy of the derived populations is typically better than 1\%. Further details of the method are given in CHIANTI Technical Report No.~16.

A set of lookup tables is provided for \emph{Solarsoft} users in the CHIANTI package, computed over the density range $\log\,N_{\rm e}=7$ to 13. For other users tables for all ions can be generated with the routine \verb|ch_lookup_all_ions|.

A /lookup keyword has been added to key IDL routines such as \verb|dens_plotter|, \verb|temp_plotter| and \verb|ch_synthetic| to implement the lookup tables. In addition the new routines \verb|ch_lookup_emiss| and \verb|ch_lookup_gofnt| yield emissivities and contribution functions from the lookup tables. A key feature of the software is that the CHIANTI version number is checked to ensure the lookup tables come from the latest version of the database.

{Finally,   we have also introduced other ways to speed-up the calculations.  One is to store the
  CE rates as compressed FITS binary tables, which are faster to read than the ascii files (which are still
  kept in the database). We have also added the option to use the LAPACK version of the {\sc invert} program
  when calculating the populations from the rate matrix. 
}

\section{Improved recombination rates and ionization equilibria}

\begin{figure}[t]
\centering
\includegraphics[width=1.08\hsize, angle=0]{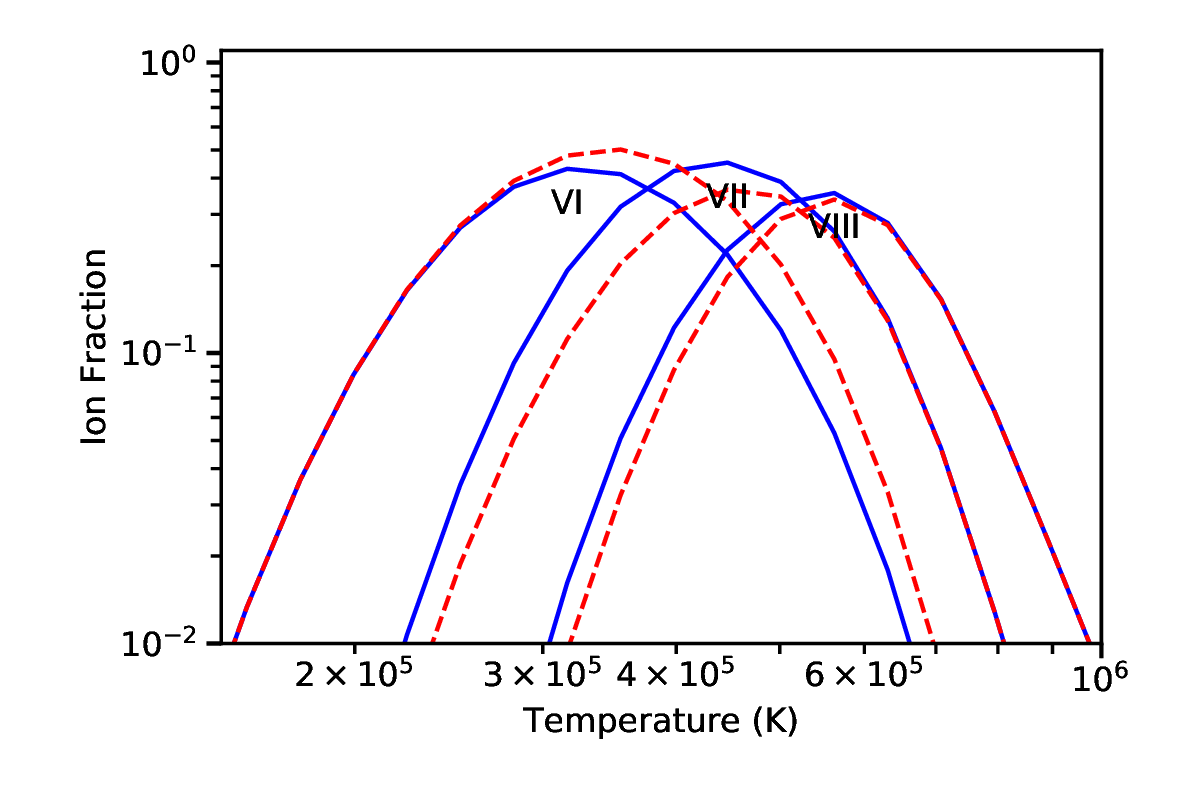}
\caption{A comparison of the new ionization equilibrium (blue line) to that of version 9 (red dashed line) for several stages of calcium}
\label{fig:ca_ioneq}
\end{figure}

New calculations of the radiative and dielectronic recombination rates for silicon-like ions recombining to phosphorus-like ions have been presented by \citet{kaur_dr_siseq}.  These have been incorporated into CHIANTI together with a new calculation of the ionization equilibrium.  They do not represent large changes in the ionization equilibrium.  For the case of calcium, the changes between the Version 9 and the new Version 10 ionization can be seen in Fig. \ref{fig:ca_ioneq}. {The temperatures of maximum ionization equilibrium for \ion{Ca}{vi} is 3.2 $\times$ 10$^5$, for \ion{Ca}{vii} is 4.5 $\times$ 10$^5$ and for \ion{Ca}{viii} is 5.6 $\times$ 10$^6$ K for the new version dataset.}

\section{Dielectronic recombination (DR)}

\subsection{Dielectronic recombination into helium-like ions}

Recently, a large-scale collisional-radiative model for helium
emission in the solar corona was developed \citep{delzanna_etal:2020_he}.
Whilst benchmarking the various atomic rates, some inconsistencies in the current CHIANTI model for neutral helium were found. However, the present model has not been 
modified as it is very limited in terms of physical processes included.
The main error that was found was in the 
dielectronic recombination (DR) into neutral helium, as calculated by N.R.Badnell.
New DR rates were recalculated by \cite{delzanna_etal:2020_he}.
The new rates are about a factor of two higher and have now been included
in CHIANTI. The error was also present (albeit not as large) for the
corresponding Li, Be, and B ions, which have also been updated in the UK APAP
repository and in the present CHIANTI version.

\subsection{Suppression of dielectronic recombination rates}

CHIANTI uses total DR rate coefficients in the calculation of the equilibrium ionization balance, and the so-called "zero-density" rates are used, meaning there is no density sensitivity. It is known, however, that DR rates become suppressed at high densities as the increased electron collision rates de-populate the high-lying levels that are important in the DR process.
\citet{summers74} modeled this process, and \cite{nikolic18} provided a fitting formula that is applicable to all ions in the CHIANTI database. With CHIANTI 10 we have coded the \cite{nikolic18} suppression factor formula and the IDL routine \verb|ch_dr_suppress| returns the DR rate coefficient for any ion for a specified electron number density ($N_{\rm e}$) or electron pressure ($P$). If the latter is specified, then the density is set to $P/T$, where $T$ is the temperature.

A consequence of the DR rates is that the equilibrium ionization balance table (stored in a CHIANTI "ioneq" file) becomes density dependent. The IDL routine \verb|make_ioneq_all| generates the ioneq file and it now takes the optional inputs "density" and "pressure".
The default CHIANTI ioneq file that is distributed with the database (named "chianti.ioneq") continues to be the zero-density table.

{We caution that the suppression factor formula is by itself an approximation
to effective recombination coefficients obtained with a simplified treatment, } 
  and that there are other sources of density sensitivity to the ionization equilibrium equations that are not modeled. In particular that due to ionization/recombination out of metastable levels, as recently discussed for carbon and oxygen ions by 
\citet{dufresne_delzanna:2019} and \citet{dufresne_delzanna:2020}.

Further details are given in CHIANTI Technical Report No.~17, and examples of the use of the new ioneq files are given in \citet{2018ApJ...857....5Y} and \citet{2018ApJ...855...15Y}.

\section{Conclusions}

As described in a recent review on the status of the atomic data and modelling available within CHIANTI \citep{delzanna_young:2020}, the present version is 
a significant step forward in terms of completeness of atomic data for 
astrophysical ions. Significant improvements are provided for several ions for which little or no data were available. The most important improvements are for the ions of the Be-like sequence, as we have shown with one example. For those cases where accurate atomic data were available in the previous version, the recent large-scale calculations (scattering and radiative) often produce  line emissivities for the strongest lines that are reasonably close (within 10-20\%) to those obtained from the previous data, providing confidence in the atomic data and also a measure on the overall uncertainties.
As briefly mentioned here and also pointed out in \cite{delzanna_young:2020}, several improvements on the modelling side are however still {required within the CHIANTI 
package. The inclusion of more physical processes such as photoionization, charge exchange, collisional ionization from excited states will be 
considered in future releases.}

\acknowledgments
 GDZ acknowledges support from STFC (UK) via the consolidated grants 
to the atomic astrophysics group (AAG) at DAMTP, University of Cambridge (ST/P000665/1. and ST/T000481/1). PRY and KPD acknowledge support from the NASA Heliophysics Data Environment Enhancements program.
The UK APAP network calculations have been funded by STFC via 
the  University of Strathclyde (grants No.
 PP/E001254/1 and ST/J000892/1., and ST/R000743/1).
 
We would like to thank all our colleagues
who provided data in electronic format or pointed out issues with the database. In particular, we thank Kai Wang, Christophe Morisset, John Raymond, 
Claudio Mendoza, Roger Dufresne, Nigel Badnell and Peter Storey.
We also thank the referees for useful suggestions to improve the paper.
EL acknowledges support from NSF grants AGS 1408789, 1460170, and NASA grants NNX16AH01G,
NNX17AD37G and 80NSSC18K0645.


\bibliography{chianti_v10}{}
\bibliographystyle{aasjournal}





\end{document}